\documentclass[fleqn,usenatbib,useAMS]{mnras}

\DeclareRobustCommand{\VAN}[3]{#2}
\let\VANthebibliography\thebibliography
\def\thebibliography{\DeclareRobustCommand{\VAN}[3]{##3}\VANthebibliography}

\usepackage{graphicx}	
\usepackage{amsmath}	
\usepackage{amssymb}	
\usepackage{multicol}  
\usepackage{bm}		
\usepackage{pdflscape}	

\newcommand\xmm{{\it XMM-Newton}}

\newcommand\astrosat{{\it AstroSat}}
\newcommand\nustar{{\it NuSTAR}}

\newcommand\swift{{\it Swift}}

\usepackage{threeparttable}
\usepackage{array}
\usepackage[T1]{fontenc}
\usepackage{mathtools}
\usepackage{nccmath}
\usepackage[english]{babel}%
\usepackage{framed}
\usepackage[normalem]{ulem}
\usepackage{xcolor}
\usepackage{braket}
\usepackage{amsfonts}
\usepackage{enumerate}
\usepackage[utf8]{inputenc}
\newcolumntype{P}[1]{>{\centering\arraybackslash}p{#1}}
\usepackage[T1]{fontenc}
\usepackage{mathtools}
\usepackage{nccmath}
\usepackage[english]{babel}%
\usepackage{framed}
\usepackage[normalem]{ulem}
\usepackage{xcolor}
\usepackage{braket}
\usepackage{amsfonts}
\usepackage{enumerate}
\usepackage[utf8]{inputenc}
\usepackage{hyperref}
\usepackage{subcaption}
\usepackage{placeins}
\usepackage{floatrow}
\usepackage{booktabs}
\usepackage{microtype}
\usepackage{color}
\usepackage{multicol}
\usepackage{multirow}
\usepackage{floatrow}
\usepackage{lipsum}
\usepackage{verbatim}
\usepackage{derivative}
\usepackage{epstopdf}
\usepackage{rotating}
\usepackage{fancybox}
\usepackage{pstricks}
\usepackage{pst-plot}
\usepackage{pstricks-add}
\usepackage{caption}
\usepackage{tabularx}
\MakeRobust{\overrightarrow}
\usepackage{pdfpages}

\title{Detection of X--ray/UV delay in NGC~4051 using \textit{AstroSat} observations}

\author[K. Kumari et al.]{Kavita Kumari$^{1}$\thanks{E-mail: kavitak@iucaa.in},
G. C. Dewangan$^{1}$,
I. E. Papadakis$^{2,3}$,
K. P. Singh$^{4}$
\\
$^{1}$Inter-University Center for Astronomy and Astrophysics, Pune 411007, India\\
$^{2}$Department of Physics and Institute of Theoretical and Computational Physics, University of Crete, 71003 Heraklion, Greece\\
$^{3}$Institute of Astrophysics, FORTH, GR-71110 Heraklion, Greece\\
$^{4}$IISER Mohali, Knowledge City, Sector 81, Manauli PO, SAS Nagar, Punjab 140306, India\\
}

\date{Accepted XXX. Received YYY; in original form ZZZ}

\pubyear{2023}

\begin{document}
\label{firstpage}
\pagerange{\pageref{firstpage}--\pageref{lastpage}}
\maketitle{}

\begin{abstract}
We study the connection between the variations in the far ultra--violet (FUV), near ultra--violet (NUV) and X--ray band emission from NGC~4051 using 4--days long \astrosat{} observations performed during 5--9 June 2016. NGC~4051 showed rapid variability in all three bands with the strongest variability amplitude in the X--ray band ($F_{var} \sim 37\%$) and much weaker variability in the UV bands ($F_{var} \sim 3 - 5\%$). Cross-correlation analysis performed using Interpolated cross--correlation Functions (ICCF) and Discrete cross--correlation Functions (DCF) revealed a robust correlation ($\sim 0.75$) between the UV and X--ray light curves. The variations in the X--ray band are found to lead those in the FUV and NUV bands by $\sim 7.4{\rm~ks}$ and $\sim 24.2{\rm~ks}$, respectively. The UV lags favour the thermal disc reprocessing model.  The FUV and NUV bands are strongly correlated ($\sim 0.9$) and the variations in the FUV band lead those in the NUV band by $\sim 13{\rm~ks}$.  
Comparison of the UV lags found using the \astrosat{} observations with those reported earlier and the theoretical model for thermal reverberation time--lag suggests a possible change in either the geometry of the accretion disc/corona or the height of the corona.

\end{abstract}




\section{Introduction}

Active Galactic Nuclei (AGNs) are powered by an accreting supermassive black hole (SMBH) at the centre of the galaxy \citep{lynden_1969Nature}, surrounded by an accretion disc. 
Variability on different timescales in ultra--violet (UV) and X--ray energy bands is a ubiquitous phenomenon of AGN, with X--rays being more variable than UV. The UV emission is expected to be the thermal emission from the disc which depends on the black hole mass ($M$) and the accretion rate \citep{Shakura_1973A&A}. 
Evidently, the X--ray photons are generated from the inverse Compton scattering of the thermal UV photons by the hot ($kT_e \sim$ 100 keV) and dynamic electrons in a compact region near SMBH known as \textit{corona}. The exact geometry of the corona is, however,  speculative. 
In principle, if the UV and X--ray emission processes are coupled, the study of correlations between X--rays and UV variations in AGN can reveal the causal connections and provide information about the disc-corona geometry.

\par The strength of correlation, time--lag value and UV/X--ray lead (or lag) depends on disc-corona configuration and the dominant processes involved \citep{Cackett_2007MNRAS, Kammoun_2021ApJ, Panagiotou2022ApJ}. The X--ray lead over UV photons by hours to days has been found to be consistent with the light crossing time--scale between accretion disc and corona and can be explained by the X--ray thermal reprocessing \citep{Kammoun_2021ApJ, Kammoun_2021MNRAS} while X--ray lagging UV has also been observed in few sources which support Thermal Comptonization model (e.g. \citealt{Sunyaev1979Nature, Nandra_1998ApJ, Arevalo_2005AA, Adegoke_2019ApJ, Pahari_2020MNRAS, Kumari2023MNRAS}). The study by \cite{Shemmer2003MNRAS} observed that a part of the optical flux exhibits variations that occur in advance of the X--ray flux by  $\sim2.5$~days. Such a delay, if real, could be indicative of fluctuations in the accretion rate propagating inwards
\citep{Lyubarskii_1997MNRAS,  Arevalo_2006MNRAS}. On the other hand, some AGNs do exhibit correlated X-ray/UV emission but without significant delays (e.g. \citealt{Breedt_2009MNRAS}). The varying lag behaviour observed from different studies suggests that there are multiple factors at play and the relationship between X--ray and UV emissions is not straightforward.

NGC~4051 is a low--luminosity Narrow Line Seyfert 1 (NLS1) galaxy located at a redshift of $z = 0.00234$.
This source is known for high variability in X--ray band. Various attempts have been made to find the X--ray and UV/optical correlation with differing and inconclusive results. One of the first studies was done by \cite{Done1990MNRAS} using 3 days-long simultaneous infrared (IR), optical, UV and X--ray observations. They reported that the X--rays vary by a factor of up to 2 on short timescales but there is no significant variation in longer wavelengths. 
Using simultaneous \textit{RXTE} and \textit{EUVE} observations, \cite{Uttley2000MNRAS} found a strong correlation between X-ray and extreme UV (EUV) band variations with a time lag between 0--20 ks. However, their X--ray light curve is not well sampled and lag is not well constrained. \cite{Mason2002} and \cite{Smith2007MNRAS} used $\sim1.5$ days long \xmm{} observation and found that the UV (2910~\AA) may lag X--rays by $\sim0.2$ days. 
\cite{Breedt2010MNRAS} utilised more than 12 years of X--ray and optical data and found that optical variations were lagging behind X--rays by $1.2^{+1.0}_{-0.3}$ days with significant correlation. 
\cite{Alston2013MNRAS} analysed the \xmm{} and \swift{} observations performed in 2009 and detected intrinsic variability in only 4 out of 15 Optical Monitor (OM) observations. The lack of intrinsic variability in UV light curves was most likely due to large systematic errors. Using the 4 good observations they obtained an average CCF plot and found a broad peak near $\sim3$ ks with only a moderate correlation of $\sim0.5$.

With the above-mentioned conflicting and inconclusive results, it is difficult to establish a coherent understanding of how different emission regions within NGC~4051 are related or interact with each other. Such studies demand high-quality, long and simultaneous X--ray and UV data.  
Here, we investigate the X-ray/UV correlation and time--lag using $\sim4$ days long, simultaneous \astrosat{} observation and try to interpret our results. In section~\ref{sec2:data}, we discuss our observation and data reduction process. In section~\ref{sec3:analysis}, we explain the cross--correlation analysis followed by discussion and conclusions in sections~\ref{sec4:discussion} and \ref{sec5:conclusion}, respectively.


\section{Data Reduction}\label{sec2:data}


\subsection{UVIT data reduction}\label{2.1}

We have used the \astrosat{} observation of NGC~4051 performed during 5--9 June 2016  (Obs.~ID:~G05\_248T01\_9000000486) with the Ultra-Violet Imaging Telescope (UVIT; \citealt{Tandon_2017AJ, Tandon_2020AJ}) as the primary instrument in photon counting (PC) mode. The UVIT is capable of observing the source in different channels -- Far Ultra-Violet (FUV, 1300--1800 \AA), Near Ultra-Violet (NUV, 2000--3000 \AA) and Visible (VIS, 3200--5500 \AA), however, the images taken in VIS channel are only used for tracking the drift. The FUV and NUV channels are equipped with multiple filters for narrower-band imaging. We used the broadband UVIT filters FUV/Silica (F172M; $\lambda_{mean}=1717$~\AA, $\Delta\lambda = 125$~\AA) and NUV/NUVB15 (N219M; $\lambda_{mean}=2196$~\AA, $\Delta\lambda = 270$~\AA) in full--frame. 

\begin{figure*}[!ht]

\centering
\begin{subfigure}{0.31\textwidth}
\includegraphics[width=1.05\textwidth]{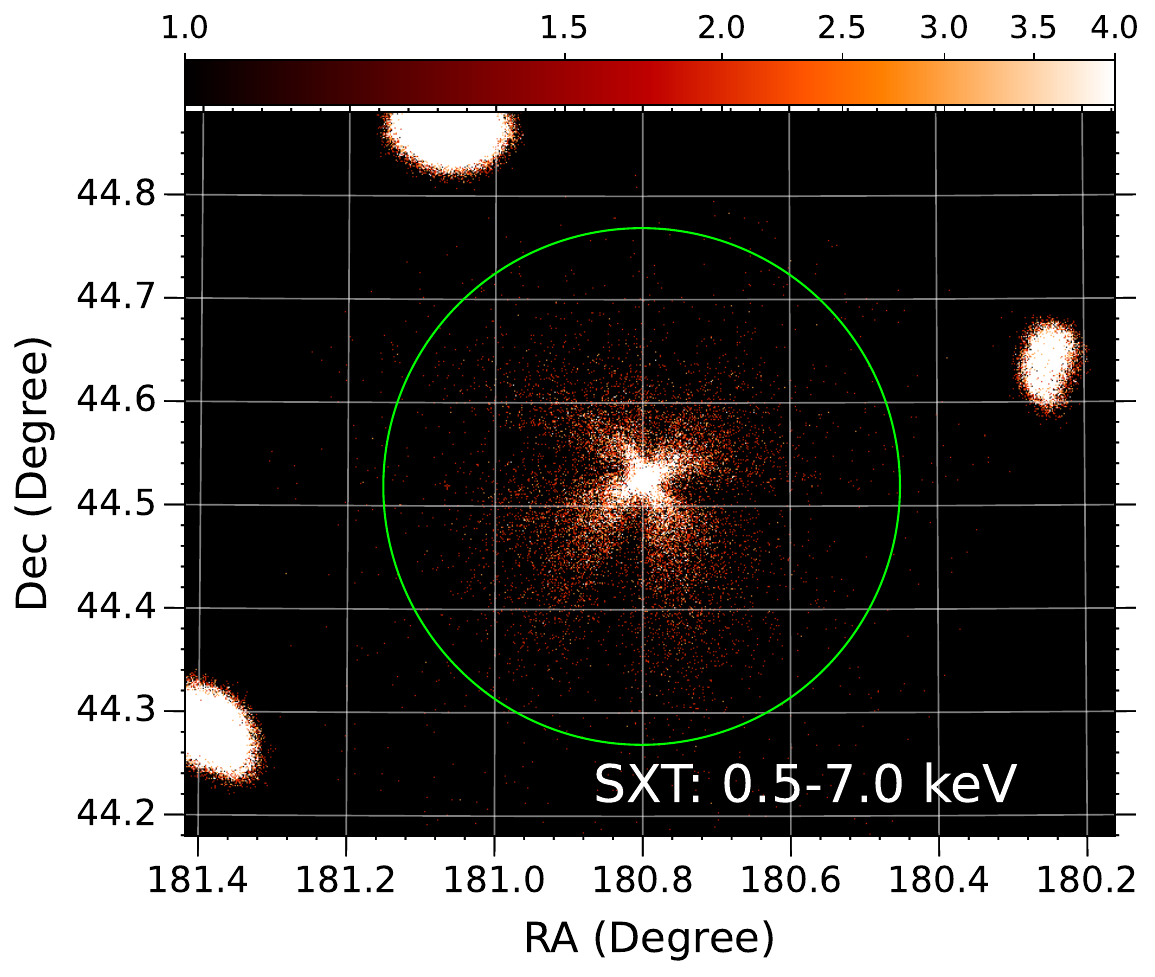}
\caption{}
\label{sxt_image}
\end{subfigure}
\hfill
\begin{subfigure}{0.31\textwidth}
\includegraphics[width=1.08\textwidth]{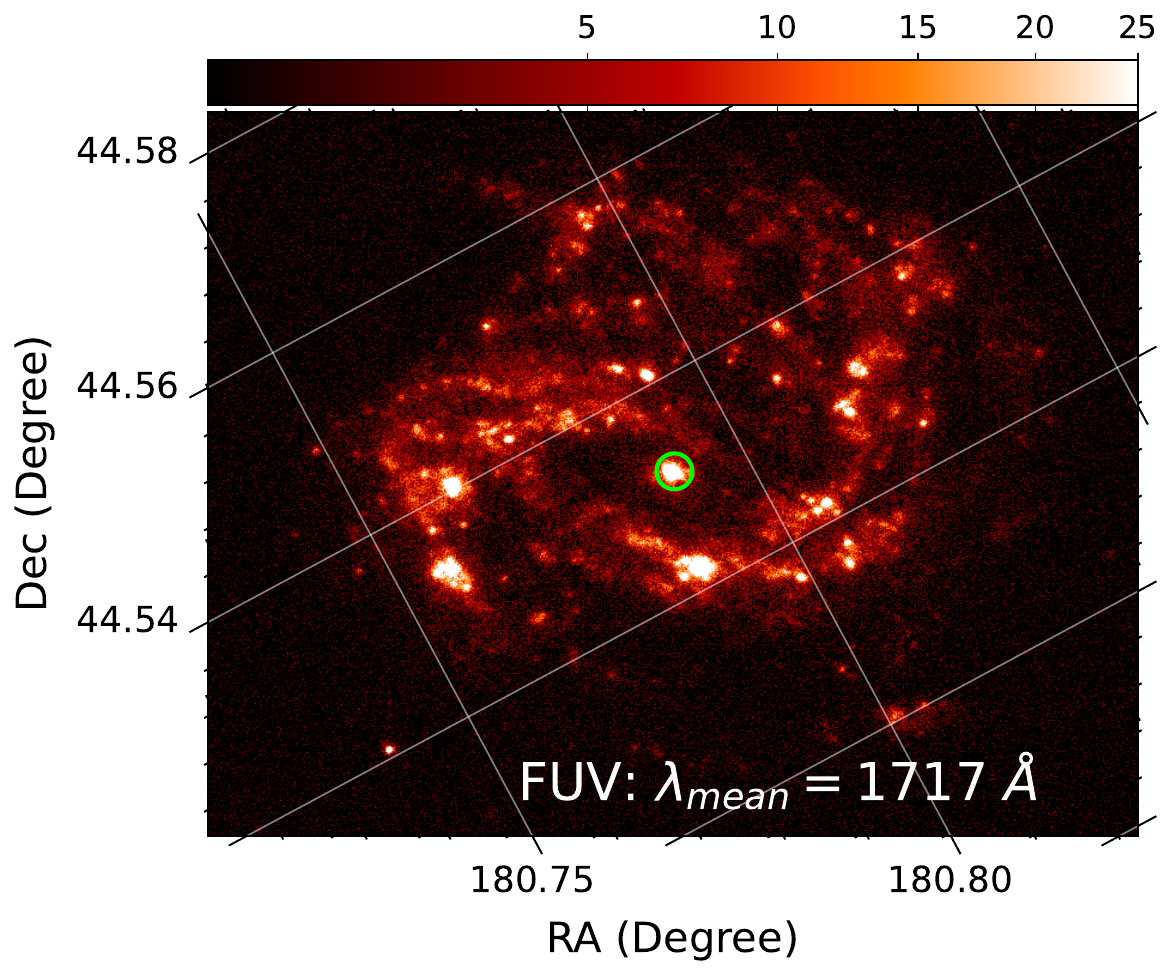}
\caption{}
\label{fuv_image}
\end{subfigure}
\hfill
\begin{subfigure}{0.31\textwidth}
\includegraphics[width=1.08\textwidth]{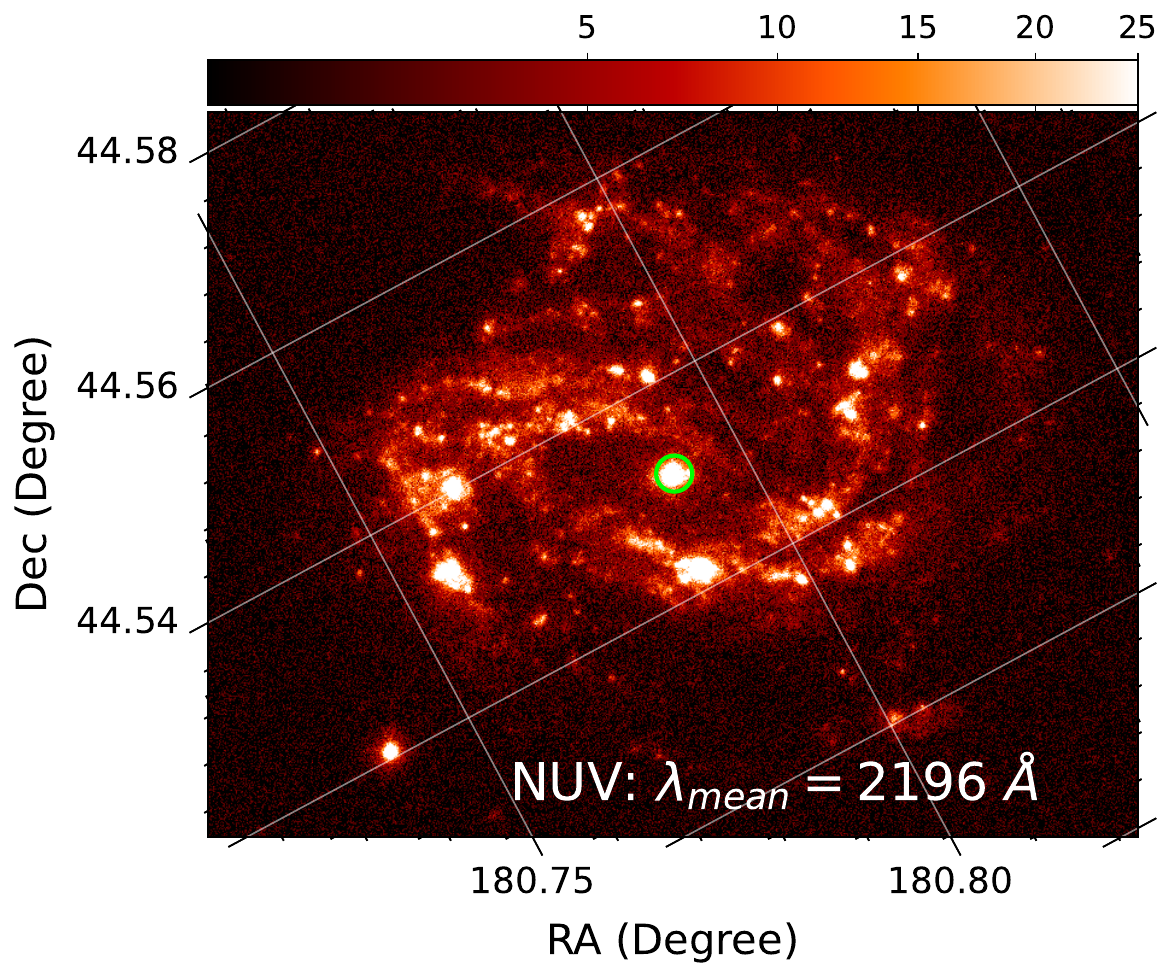}
\caption{}
\label{nuv_image}
\end{subfigure}
\caption{\astrosat{} images of NGC~4051 in different energy bands: (a) SXT (0.5 -7.0 keV), (b) FUV/Silica (F172M; $\lambda_{mean}=1717$~\AA) and (c) NUV/NUVB15 (N219M; $\lambda_{mean}=2196$~\AA). The circular source extraction  regions (green) measure 15 arcmins for SXT and 0.1 arcmins (6.25 arcsecs) for UVIT images. For plotting Right Ascension (RA) and Declination (Dec) coordinates, we have utilised Python--based package APLpy \citep{aplpy2012ascl}.} 

\label{images}
\end{figure*}

We obtained the Level1 (L1) UVIT data from the \astrosat{} data archive\footnote{\url{https://astrobrowse.issdc.gov.in/astro_archive/archive/Home.jsp}} and processed the data with the \textsc{CCDLAB} UVIT Pipeline \citep{Postma_2017PASP} to get the final orbitwise images. Please refer to \cite{Kumari2023MNRAS} for the details of data reduction using CCDLAB. 
We have developed the {\sc UVITTools} package in the Julia language \citep{dewangan_2021JApA} to get the event list file from an individual orbitwise image of AstroSat/UVIT data processed with CCDLAB. To obtain the event list we use the time calibration table (.tct~file). Finally, using the same package, we merged all the event lists from different orbits for the \textit{``merged\_level2.evt"} which can be used to extract the light curve using the HEASoft\footnote{\url{https://heasarc.gsfc.nasa.gov/lheasoft/
}} tool \textsc{XSELECT} (version 2.5a). 
We used the circular regions with radii 15~pixels or 6.25~arcsecs, centred at the source position to extract the source counts  and a circular region of 60~pixels in a source--free region for the background counts with a time-bin size of 1 sec. Then we used the standard HEASoft tool, ``lcmath", to get the net background-subtracted light curve of 1 sec bin--size (see Fig.~\ref{fuv_image} and \ref{nuv_image}). 
 
We rebin our light curve using the same technique as explained in \cite{Kumari2023MNRAS}, for the same reasons. Initially, we tried to rebin the light curves with smaller bin--sizes (less than $\sim97$ minutes which is the orbital time period of the \astrosat{}) but that led to some fallacious data points due to the large background. We, therefore, rebinned the FUV and NUV light curves extracted with 1~sec bins to a time bin--size of 5846~sec (or $\sim97$ minutes) for our study shown in this paper.


\begin{table}[!ht]
\caption{Details of the \astrosat{} observation of NGC~4051 for Observation ID: G05\_248T01\_9000000486.}
\renewcommand{\arraystretch}{2}
\begin{tabular}{P{0.11\linewidth}P{0.2\linewidth}P{0.13\linewidth}P{0.11\linewidth}P{0.17\linewidth} }\\
\hline
\hline
\textbf{Obs. date} & 
\textbf{Energy band/Filter} & \textbf{Mean count rate (c/s)} &\textbf{Exp. time (ks)} &\textbf{$F_{rms}$} (\%)\\
\hline

  5-9 June 2016 & SXT/(0.5-7.0 keV) & 0.73  & 69.5 &$38.6\pm0.66$ \\
  
& FUV/F172M & 1.03   &  35.9  & $4.67\pm0.68$ \\

 & NUV/N219M & 1.93  & 36.8   & $3.51\pm0.51$ \\

\hline
\hline
\end{tabular}
\label{data}
\end{table}


\begin{figure}[!ht]
\includegraphics[width=1\columnwidth]{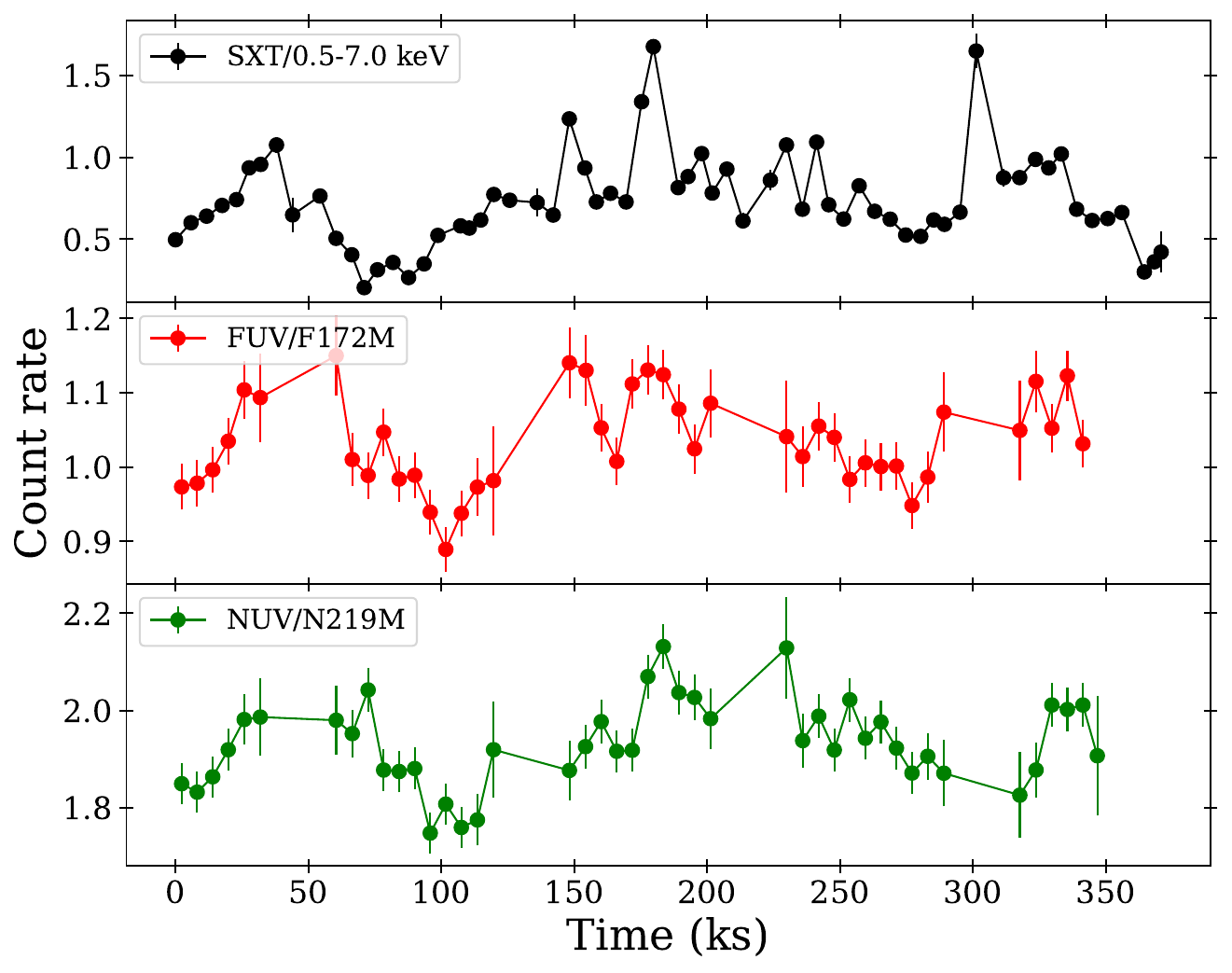}
\caption{Light curves of NGC~4051 in different energy bands: SXT (top panel), FUV/F172M (middle panel) and NUV/N219M (bottom panel).}
\label{lc}
\end{figure}


\subsection{SXT data reduction }\label{2.2}

For the X--ray light curve, we have utilised the simultaneous data of Soft X-ray Telescope (SXT; \citealt{singh_2016SPIE, singh_2017JApA}). The SXT is an imaging instrument with a limited spatial resolution (FWHM\footnote{FWHM: Full Width at Half Maximum}$\sim 2{\rm~arcmins}$, half-power diameter $\sim 10{\rm~arcmins}$), and a circular field of view (FOV) of $\sim 40{\rm~arcmins}$ diameter.
Due to the large point spread function (PSF), there is hardly any source-free area in the CCD detector to get the background count, so, observations of the blank sky are used for background correction (SkyBkg\_com-
b\_EL3p5\_Cl\_Rd16p0\_v01.pha) provided by the SXT POC.

We downloaded the orbit-wise L1 SXT data on NGC~4051 from the \astrosat{} archive and processed using the  {\sc SXTPIPELINE} (Version:~1.4b), available at  the SXT website\footnote{\url{https://www.tifr.res.in/~astrosat_sxt/index.html}}, to obtain the orbit--wise level2 data that include the cleaned event list for each orbit. We merged
the orbit-wise event lists  using the Julia package  {\textsc SXTMerger.jl}\footnote{\url{https://github.com/gulabd/SXTMerger.jl}}, also available at the SXT website. 
We used the HEASoft tool \textsc{XSELECT} (version 2.5a) to extract the images and light curves from the merged event lists. 
We extracted the source light curves in the 0.5-7.0~keV band with 2.3775~sec (time resolution of SXT) bins using a source extraction region of 15~arcmins radius (Fig.~\ref{sxt_image}) and rebinned the light curve with bin--size of 5846~sec, similar to the UVIT case as described in \ref{2.1}.

In  Fig.~\ref{lc}, we plot the X-ray and UV light curves of NGC~4051.
We list the details of the observations including the net exposure time and mean count rates in Table~\ref{data}.


\begin{figure*}[!ht]

\centering
\begin{subfigure}{0.31\textwidth}
\includegraphics[width=1.1\textwidth]{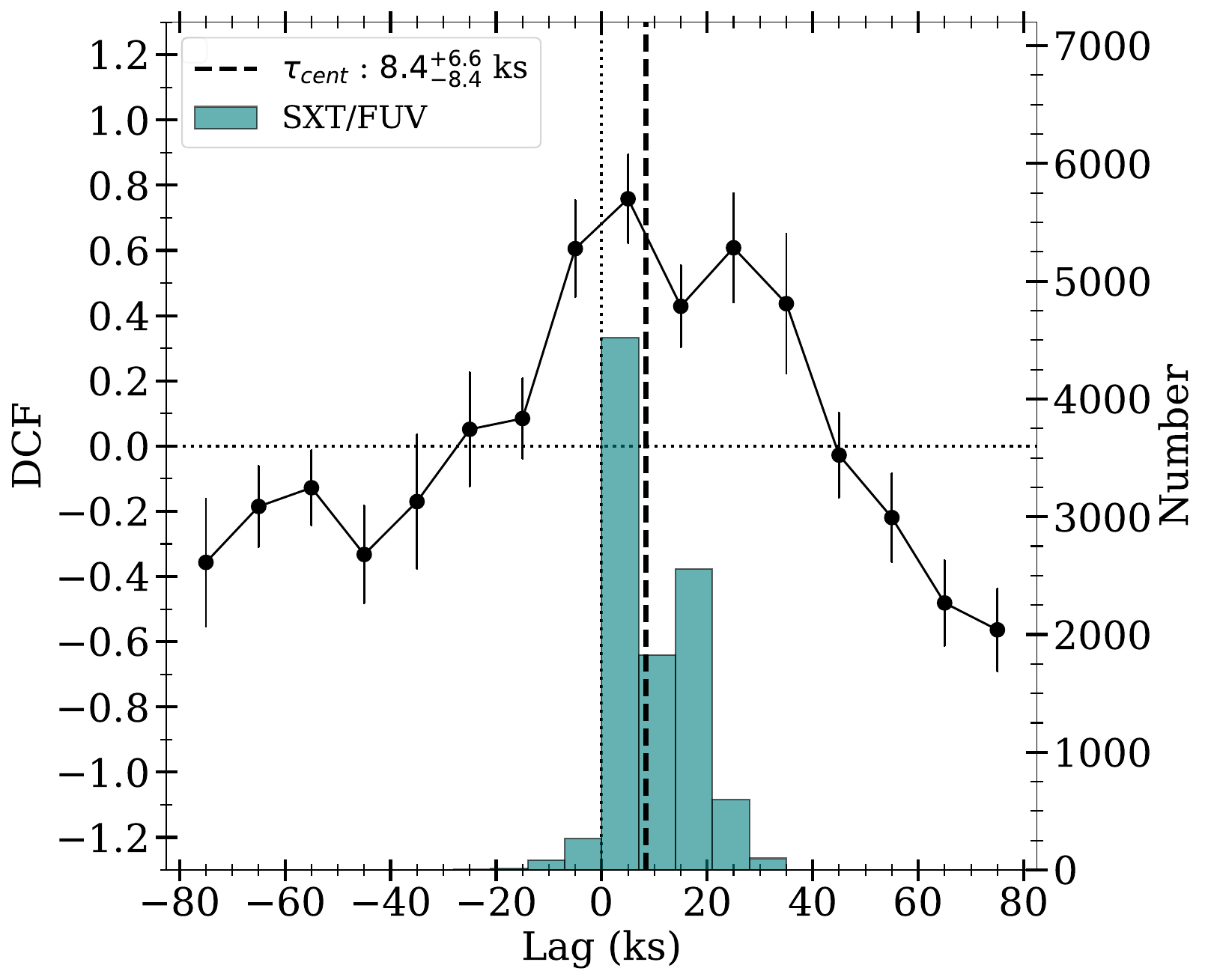}
\includegraphics[width=1.1\textwidth]{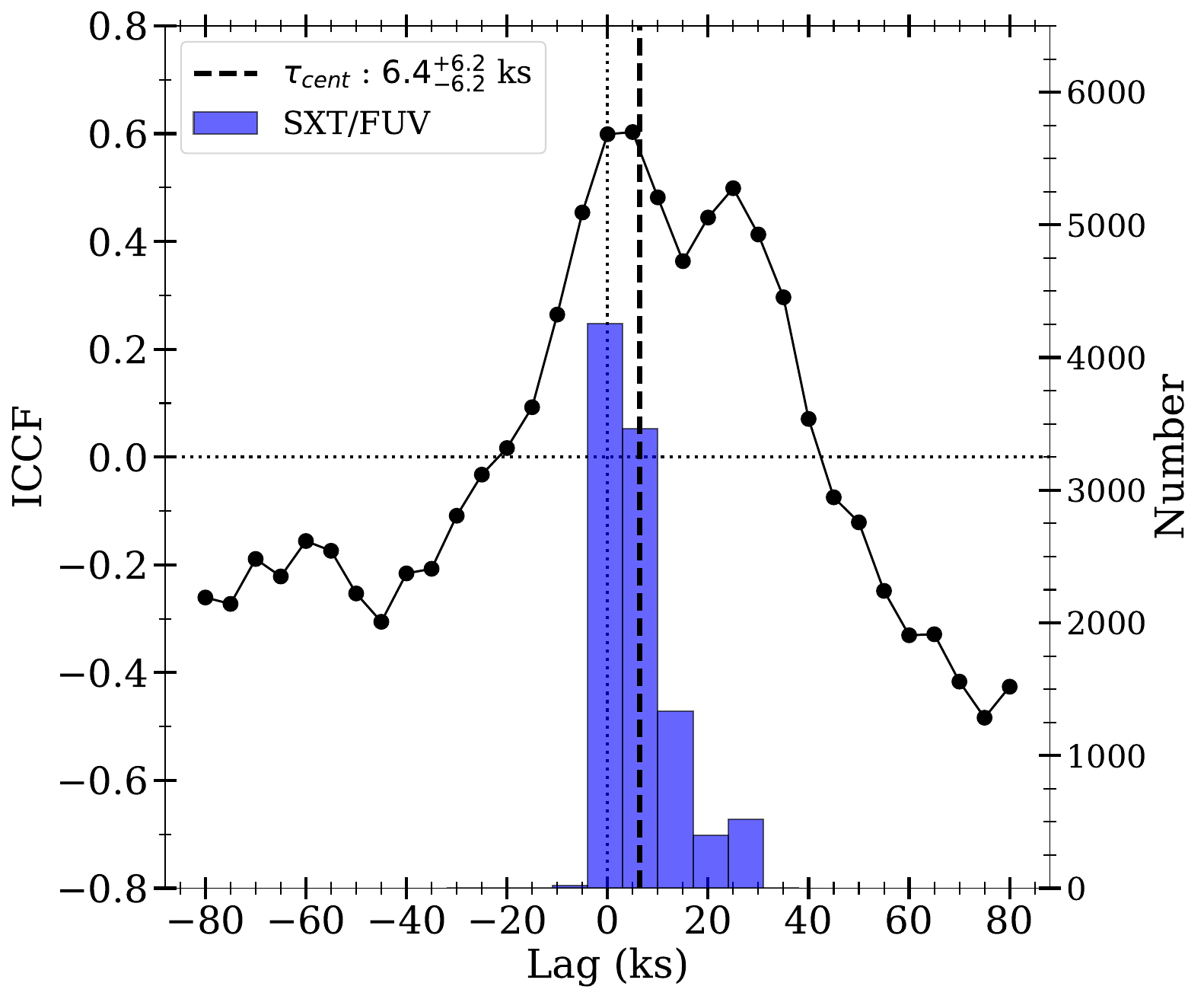}
\caption{}
\label{sxt_fuv_dcf}
\end{subfigure}
\hfill
\begin{subfigure}{0.31\textwidth}
\includegraphics[width=1.1\textwidth]{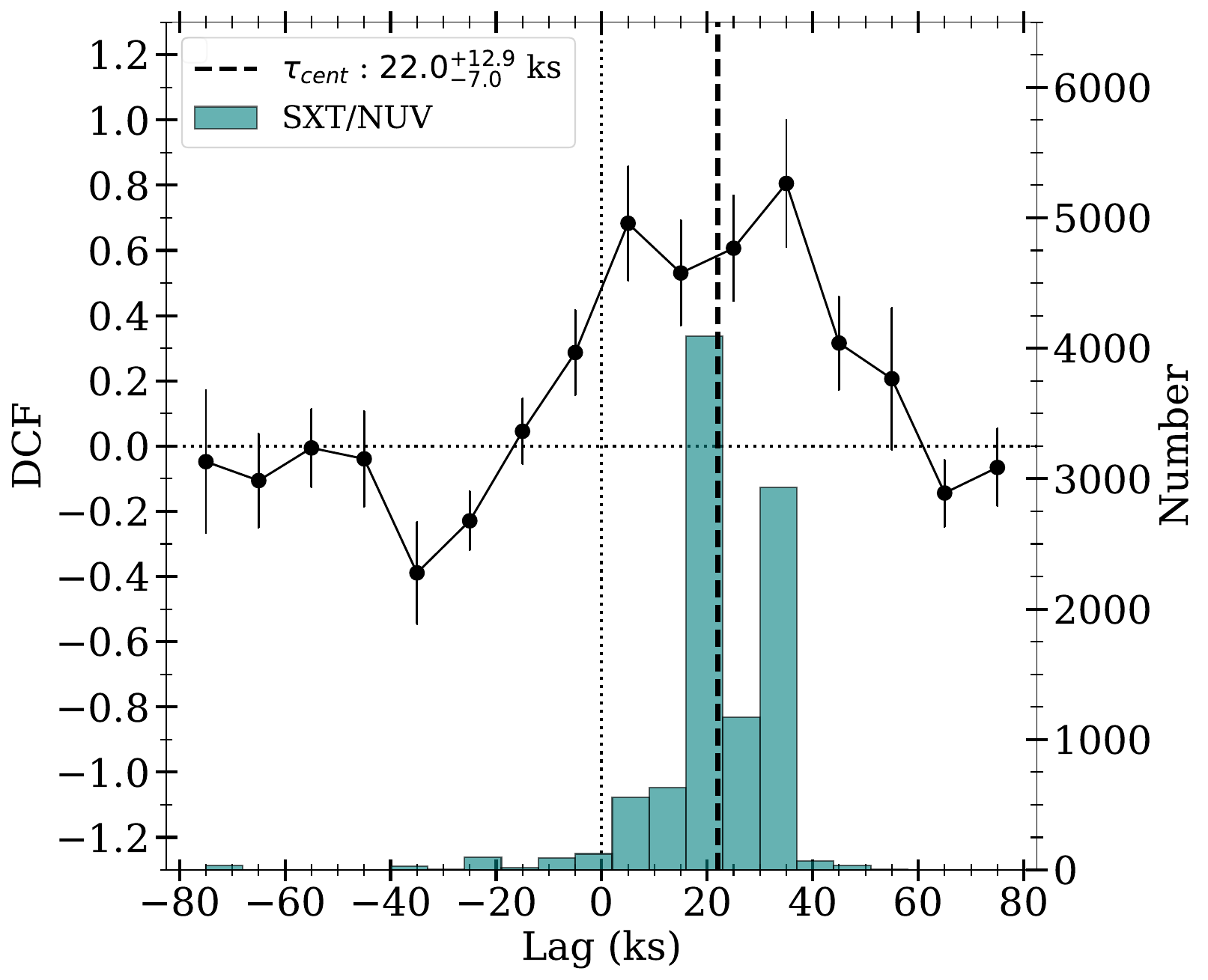}
\includegraphics[width=1.1\textwidth]{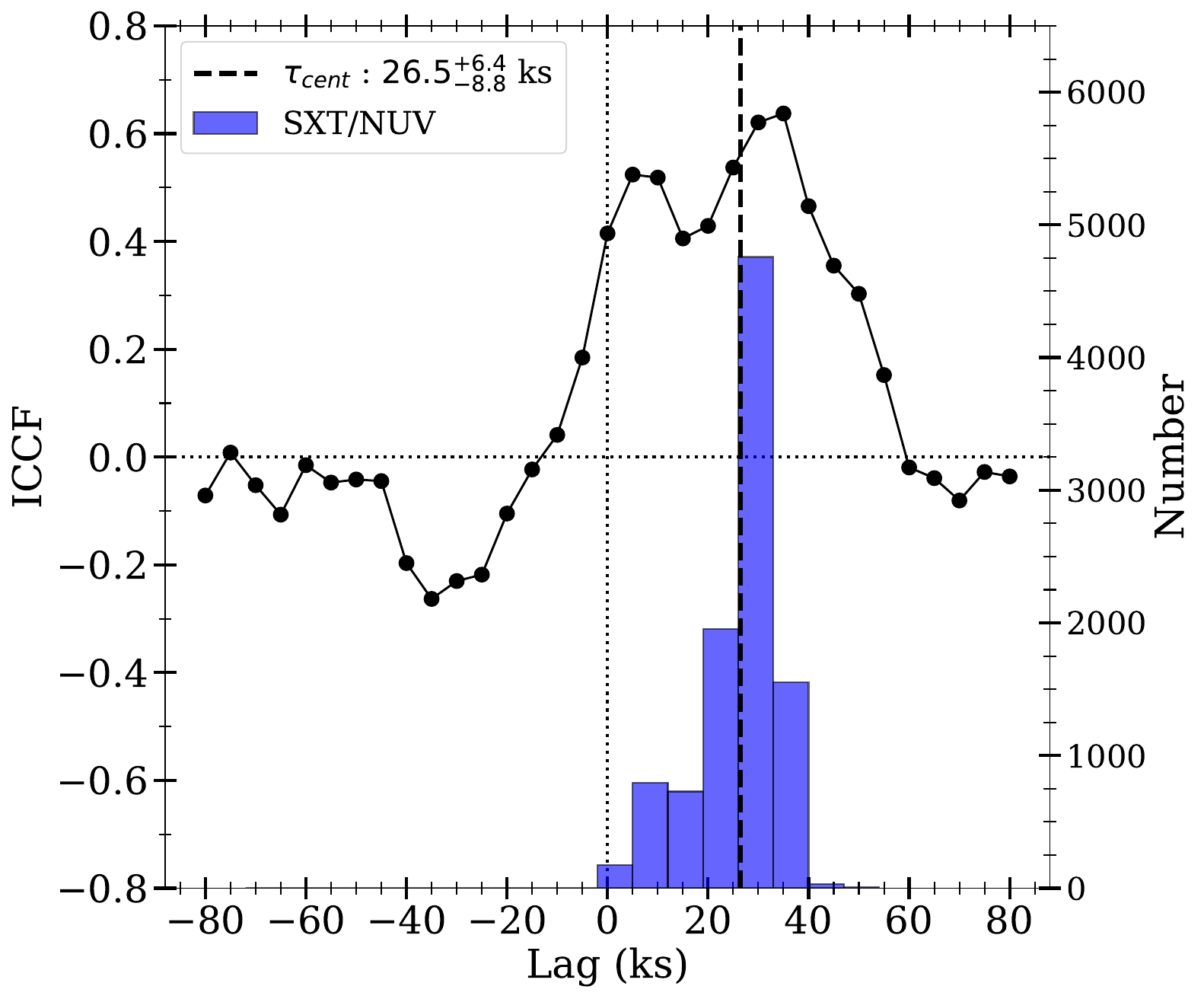}
\caption{}
\label{sxt_nuv_dcf}
\end{subfigure}
\hfill
\begin{subfigure}{0.31\textwidth}
\includegraphics[width=1.1\textwidth]{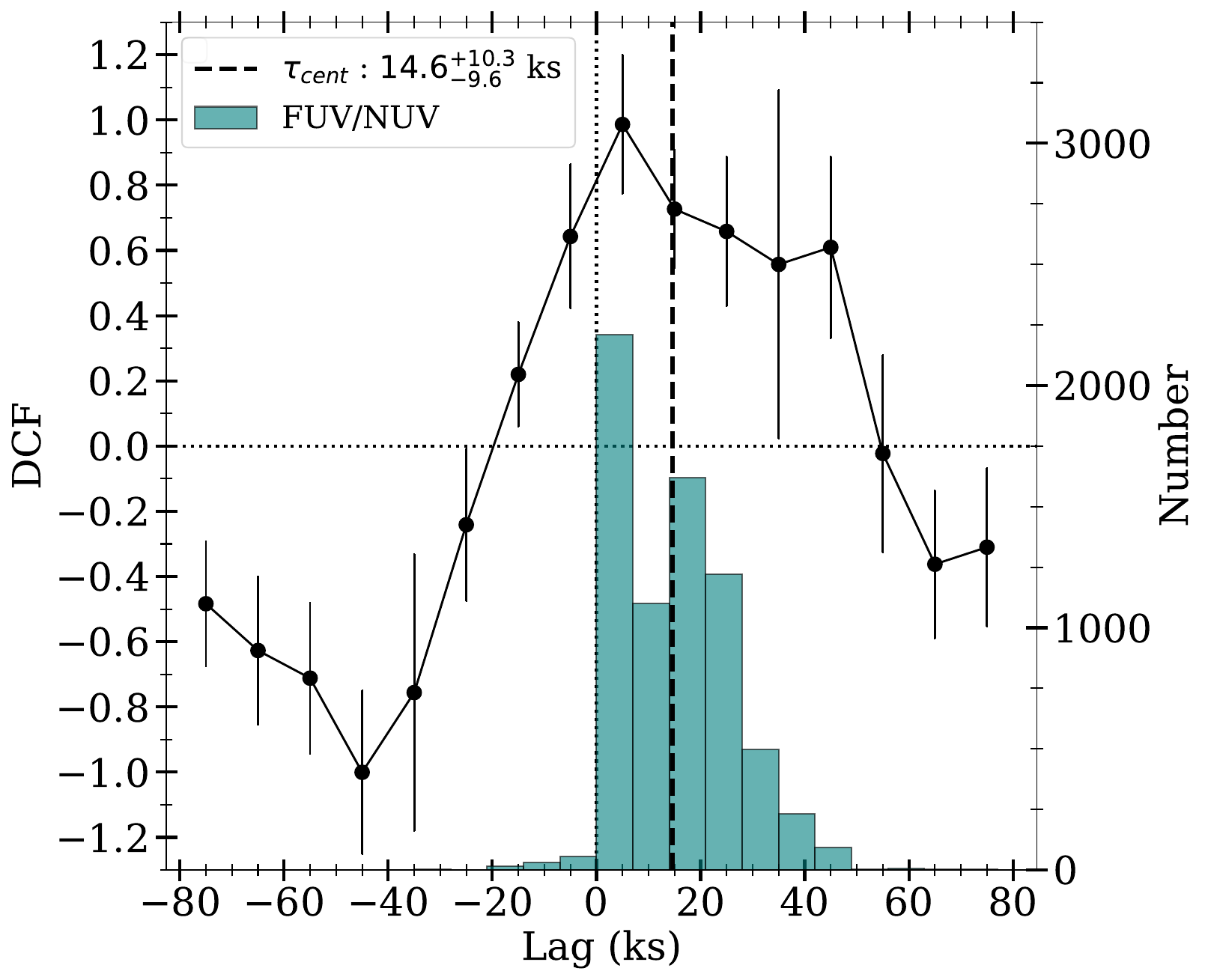}
\includegraphics[width=1.1\textwidth]{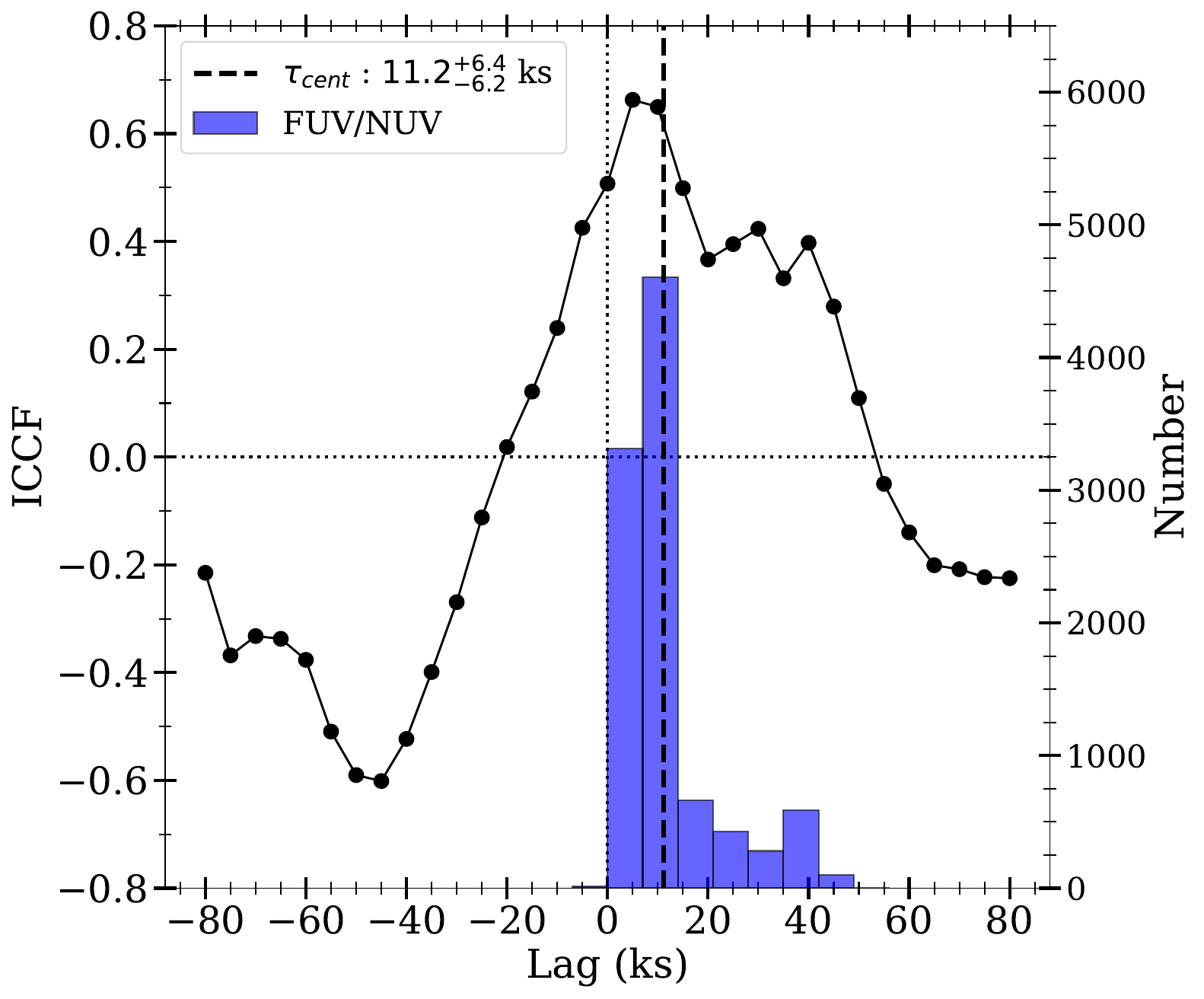}
\caption{}
\label{fuv_nuv_dcf}
\end{subfigure}
\caption{\textit{Upper panel:} The DCF (black curve) for NGC~4051  using the \astrosat\, light curves for, (a) SXT/FUV, (b) SXT/NUV and (c) FUV/NUV (left, centre and right panels, respectively).  Histograms (in teal colour) are the centroid lag ($\tau_{cent}$) distributions for 10,000 bootstrap realizations. The vertical dashed line is the mean of the  $\tau_{cent}$ distribution. \textit{Lower panel:} Same for ICCF.} 

\label{dcf_plots}
\end{figure*}


\section{Timing Analysis}\label{sec3:analysis}

\subsection{Light curves variability}

We checked the variability of the light curves shown in Fig.~\ref{lc} by fitting a constant and found $\chi^2=7457/64$ degrees of freedom (dof) 
in the SXT band and  $\chi^2=129/42$ dof, $p_{nul}=9.1\times 10^{-11}$ and $\chi^2=145/42$ dof, $p_{nul}=7.4\times 10^{-13}$ for the FUV and the NUV light curves, respectively ($p_{nul}$ is the probability of the source being constant). 
These results indicate that the source is significantly variable in all three bands.

We also calculated the fractional variability amplitude ($F_{var}$)  of the light curves using the formula as defined in \cite{Vaughan_2003MNRAS}.
The obtained $F_{var}$ results are listed in the Table \ref{data}.
We notice that the source is significantly more variable in the X--rays than in the UV band. Also, the light curve in the FUV band appears to be more variable than the NUV, however, the difference between the fractional variability amplitude in these two light curves ($F_{var, FUV}- F_{var, NUV}$) is $1.16 \pm 0.85 $\%, which is statistically not significant.

\subsection{Time--Lags using different techniques}

We computed the time--lags for the X-ray and UV light curves using the Interpolated cross--correlation Function (ICCF; \citealt{peterson_1998PASP}) and  Discrete cross--correlation Function (DCF; \citealt{edelson_1988ApJ}) techniques. For our analysis, we utilized two Python-based packages: PyCCF \citep{Sun_2018} and PyDCF\footnote{\url{https://github.com/astronomerdamo/pydcf}}. In our DCF analysis, we incorporated weighted errors to mitigate the impact of significantly large errors resulting from shorter exposure times (for details, see section 5 in \citealt{Kumari2023MNRAS}). Note that the ICCF technique does not use measurement errors in the CCF calculation. In order to obtain the lag distribution and associated uncertainties, we employed random subset selection (RSS) to generate 10,000 pairs of bootstrap realizations.
We plot our cross--correlation results in Fig.~\ref{dcf_plots}. In the upper panel, the DCF is represented by filled circles with errors, while the centroid time-lag ($\tau_{cen}$) distribution is shown in a teal--coloured histogram. Similarly, in the lower panel, ICCF results are depicted with filled circles (without errors) and histograms in blue for the SXT/FUV, SXT/NUV, and FUV/NUV cases. The ($\tau_{cen}$) is computed by averaging the time--lags corresponding to CCF values > 0.8 CCF$_{max}$. The histogram illustrates the distribution of $\tau_{cen}$ for all the RSS synthetic light curves. The vertical dashed line marks the mean of this distribution. Additionally, we calculate the centroid DCF (DCF$_{cen}$) by averaging over DCF values > 0.8 DCF$_{max}$. The mean of the $\tau_{cen}$ distribution and the respective DCF$_{cen}$ are reported in Table~\ref{results_table}. The errors provided in Table~\ref{results_table} (and throughout the paper) signify the 68 \% confidence limits for each parameter. A positive $\tau_{cen}$ means variations in UV are lagging X--ray band.

The correlation strength (DCF$_{cen}$) for FUV/NUV is $\sim0.9$ which indicates a very good correlation between the variations in the two bands. This also implies a common origin for the FUV and the NUV variations. The SXT/UV correlation strength is weaker ($\sim0.75$) but still quite large compared to the X--ray/UV correlation observed in other AGNs (see for e.g. Table 1 in \citealt{Panagiotou2022ApJ}).
The mean $\tau_{cen}$ are $\sim$~7~ks ($\sim$~0.08~days)  and $\sim$~24~ks ($\sim$~0.28~days) for the X--ray/FUV and X-ray/NUV correlations, respectively. The lag between the FUV and NUV light curves is $\sim$~13~ks ($\sim$~0.15~days).
The time--lags calculated from both techniques are consistent within uncertainty and show that the UV photons are lagging the X--rays by their respective time--lag values. 


\begin{table}[!ht]
\caption{cross--correlation results (errors correspond to 68\% confidence region for each parameter).}
\renewcommand{\arraystretch}{2}
\begin{tabular}{P{0.2\linewidth}P{0.2\linewidth}P{0.2\linewidth}P{0.2\linewidth} } \\

\hline
\hline 
& \textbf{Mean  $\mathbf{DCF_{cent}}$} & \multicolumn{2}{c}{\textbf{Mean  $\mathbf{\tau_{cent}}$ (ks)}}\\
\textbf{Filter} & & \textbf{DCF} &\textbf{ICCF} \\ 
\hline

SXT/FUV & $0.70^{+0.09}_{-0.09}$ & $8.4^{+6.6}_{-8.4}$  &  $6.4^{+6.2}_{-6.2}$ \\ 

SXT/NUV & $0.74^{+0.09}_{-0.09}$  & $22.0^{+13.0}_{-7.0}$  & $26.5^{+6.4}_{-8.8}$ \\

FUV/NUV & $0.85^{+0.09}_{-0.1}$ &$14.6^{+10.3}_{-9.6}$  & $11.2^{+6.4}_{-6.2}$ \\ 

\hline
\hline
\end{tabular}
\label{results_table}
\end{table}

\subsection{Comparison with X--ray reverberation time--lag model}\label{sec:K21}


\begin{figure*}[!ht]
\begin{subfigure}{1\textwidth}
\centering
\includegraphics[width=0.8\textwidth]{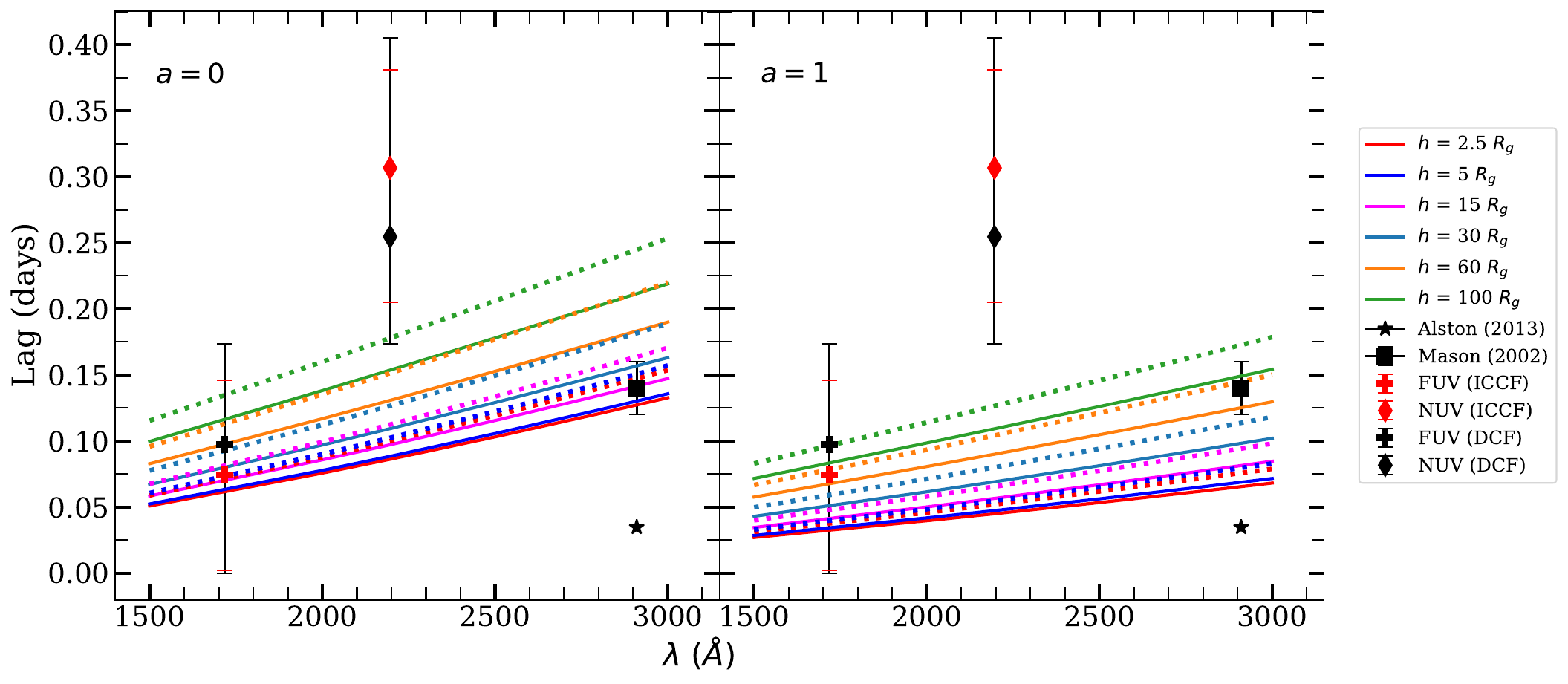}
\includegraphics[width=0.8\textwidth]{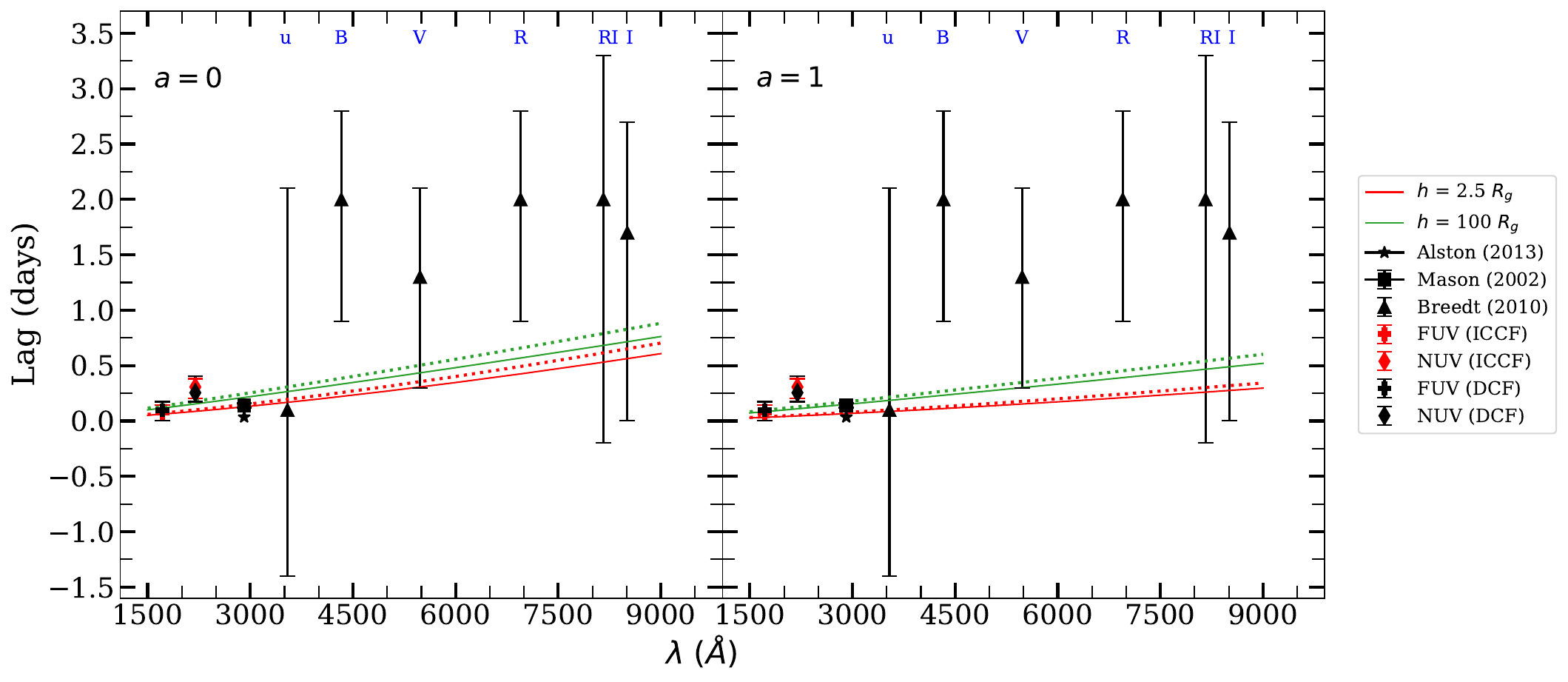}
\end{subfigure}
\caption{Comparison of observed time--lags in different bands and theoretical time--lags (solid lines) expected in the case of X--ray thermal reverberation for different corona heights (see section \ref{sec:K21}). The heights of the corona used for plotting are $h=$~2.5, 5, 15, 30, 60 and 100 $R_g$. The left and right panels are for spin 0 and 1, respectively. The dotted lines correspond to the lags for an upper limit on the BH mass used in this paper. \textit{Upper panel:} We have over--plotted our lag results for SXT/FUV (plus) and SXT/NUV (diamond) obtained using DCF (in black) and ICCF (in red) techniques and compared it with the previous results by \protect\cite{Mason2002} (square) and \protect\cite{Alston2013MNRAS} (star). \textit{Lower panel:} The model time--lag is extrapolated to higher wavelengths and also included X-ray/optical lags obtained by \protect\cite{Breedt2010MNRAS} (triangle). The model time--lag has been plotted for only two values of height, $h=$~2.5, and 100 $R_g$ for clarity.}
\label{lag_lambda}
\end{figure*}

Recently, \cite{Kammoun_2021ApJ} have computed the response function ($\Psi$) of the accretion disc and the expected time--lag, assuming a compact corona above the BH (lamp--post geometry) illuminating a standard Novikov--Thorne (NT) accretion disc (\citealt{NT1973}). They have taken into account all the special and general relativistic effects as well as the disc ionization effects in order to find X--ray/UV delay. They have also provided an analytic expression (equations (8) to (16) in their paper) for centroid time--lag ($\tau_{cen}$) as a function of wavelength ($\lambda$), the BH mass ($M_{BH}$), the absorption--corrected luminosity ($L_{X,Edd}$) in 2-10 keV band in units of Eddington luminosity,
corona height ($h$) and accretion rate ($\dot{m}$), considering BH spin ($a$) 0 and 1. \cite{Kammoun_2019ApJ, Kammoun_2021MNRAS} have demonstrated that the analytic prescription given by \cite{Kammoun_2021ApJ} can successfully model the observed time--lag spectra in seven AGNs.

The results obtained from the \astrosat{} observations, and findings from other missions along with the theoretical model can constrain the delays between X--rays and UV/optical bands. 
The black hole mass of NGC~4051 used for computing the model lag is $M_{BH} = 8.6^{+ 2.0}_{- 2.6} \times 10^5 M_\odot $, taken from the AGN BH mass databse \citep{massdatabase2015} and calculated using recent scaling factor ($f = 4.8$) given in \cite{Batiste2017}. To obtain the 2-10 keV intrinsic flux we have fitted the SXT spectrum (see Appendix \ref{spec_fitting}) and using the redshift--independent distance, $D=16.6~$ Mpc \citep{Yuan2021}, we calculated 2-10 keV luminosity $L_{X,Edd} = 0.007~L_{Edd}$. The accretion rate used in the equations is $\dot{m} = 0.2 $ \citep{Yuan2021}.
The plot of the model time--lag as a function of $\lambda$ for different heights has been shown in Fig. \ref{lag_lambda}. 
The dotted lines in Fig. \ref{lag_lambda} correspond to the lags calculated using an upper limit on the BH mass ($M_{BH} = 1.06 \times 10^6 M_\odot $).

In the upper panel of Fig. \ref{lag_lambda},  we over--plot our FUV (denoted by the plus symbol) and NUV (denoted by the diamond symbol) lag results obtained using ICCF and DCF methodologies. Additionally, we have compared our lag findings with the previous outcomes of \cite{Mason2002} (represented by a square) and \cite{Alston2013MNRAS} (represented by a star).  In the case of spin 0, the \astrosat{} time--lags could be consistent with the model predictions for large heights of corona. On the other hand, \cite{Mason2002} time--lag is consistent with a small height. If correct, this would suggest that the height in NGC~4051 varies over time. In the case of spin 1, even the \cite{Mason2002} time--lag measurement is consistent with a large corona height. However, the \astrosat{} SXT/NUV time--lag appears to be more inconsistent with the model predictions. We note that contrary to \cite{Kammoun_2021MNRAS}, here we plot the time--lags with respect to X--rays (and not with respect to the UV band). As \cite{Kammoun_2021ApJ} stated (see their equation 7 and discussion in section 4.1), the CCF between the X--ray and the UV/optical bands depends on the accretion disc response ($\Psi$) to X--rays as well as on the  X--ray autocorrelation function (ACF). The  X--ray ACF is expected to be broad in AGN, however, since the model time--lags do not incorporate the effects of X--ray ACF,  it could significantly affect the time--lags.

However, the major challenge for the model  time--lags is to fit the low lag measurement reported by \cite{Alston2013MNRAS}. The model cannot account for such a low time delay between X--rays and the UV emission at $\sim2900$~\AA~ even if the corona height is as small as 2.5 $R_g$. The discrepancy is even more pronounced if we consider that the true time--lags between X--rays and UV should be larger than the model lines plotted in Fig. \ref{lag_lambda}. Since \cite{Alston2013MNRAS} do not provide an error estimate for their lag measurement, therefore, we cannot ascertain the level of disagreement between their reported lag and the model prediction.

In the lower panel of Fig. \ref{lag_lambda}, we have extrapolated the model lag to a higher wavelength and included the centroid time--lag between X--rays and optical bands (u, B, V, R, RI, I) as reported by \cite{Breedt2010MNRAS} (shown by triangles). They have used $\sim12$ years of X-ray and optical data from different missions. We notice that their lag values in the optical bands (except the u band) are quite large (> 3 times) than the model prediction even for the corona height of 100~$R_g$. We do not expect such a large delay between X--ray and optical photons for a low--mass AGN like NGC~4051 ($M_{BH}\approx 10^6 M_{\odot}$) in the case of X--ray thermal reverberation. However, it is worth noting that, within the errors, the results obtained by \cite{Breedt2010MNRAS}  in the u, RI and I bands are consistent with the model predictions. This suggests that the discrepancies observed in the X-ray/optical time--lags reported by \cite{Breedt2010MNRAS} could potentially be attributed to the large bin--size used in their light curves for computing lag.


\section{Discussion}\label{sec4:discussion}


The study of inter-band variability and detection of time--lag requires sufficiently long, multi-band observations with good sampling. We have utilized a $\sim4$ days long \astrosat{} UV and X-ray observations with an orbital time of 97 minutes and observing efficiencies of 15\% and 25\% in the UV and X--ray bands, respectively. 
The \astrosat{} light curves of NGC~4051 in X--rays as well as in UV bands binned to orbital period show a significant variation on the timescale of < 0.5 days (see Fig.~\ref{lc}). Their fractional variability amplitude (see Table~\ref{data}) reveals that the source is more variable in X--rays, similar to what has been previously observed in other AGNs (see for example \citealt{edelson_2019ApJ, Kumari2023MNRAS}). We obtained the cross--correlation strength and time--lags between the light curves in soft X--ray, FUV and NUV bands using  ICCF and DCF techniques. 
The results listed in Table~\ref{results_table} show that the cross--correlation lags derived from both the techniques are fully consistent with each other, within the uncertainties. We report a high average correlation strength ($\sim$~0.9) between UV bands and a moderate correlation ($\sim$~0.75) between X--ray and UV bands.


We found that variations in the FUV and NUV bands significantly lag soft X-rays by $\sim 7$~ks and $\sim 24$~ks, respectively. These soft (UV) lags favour the disc reprocessing model. We also noticed a secondary peak in DCF and ICCF curves and a hint of bimodal distribution in the centroid--lag distributions (see Fig.~\ref{dcf_plots}). However, considering the data quality and gaps in the light curve, we refrain from over--interpreting our results and we only assume a broad peak around the observed lags (dashed vertical line in Fig.~\ref{dcf_plots}). 
The high correlation ($\sim 0.9$) between FUV and NUV light curves suggests a common origin of their variations. We also observed that the NUV light curve lags FUV by $\sim 13 {\rm~ks}$ which is approximately equal to the difference between the time--lag in SXT/FUV and SXT/NUV cases. This indicates that most of the detected NUV photons originate from larger disc areas as compared to FUV photons.


We have plotted the centroid time--lag as a function of wavelength for different coronal heights using the analytic prescription given by \cite{Kammoun_2021ApJ} (discussed in section \ref{sec:K21}). In the past, \cite{Kammoun_2019ApJ, Kammoun_2021MNRAS} have used these analytic equations to model the lag of a few AGNs and put possible constraints on the accretion rate and height of the corona. Our lag results using the \astrosat{}'s observation along with the previous results obtained by \cite{Mason2002}, \cite{Alston2013MNRAS} and \cite{Breedt2010MNRAS} can constrain the delay between X--ray and UV/optical bands and also the height of the corona.


We notice that the observed lag values for SXT/NUV  are slightly higher than what is predicted by the analytic equations for both cases, $a=0$ and $a=1$. This discrepancy is possible because the model time--lags do not take into account the effects of X--ray ACF (as explained in section \ref{sec:K21}).
The uncertainty in the measured values of 2-10 keV luminosity and accretion rate can also affect the model time--lags to some extent (solid lines in Fig.~\ref{lag_lambda}). For the currently used $L_{X, Edd}$ and $\dot{m}$ values, the lag for a given height can also increase with the increase in BH mass (2 or 3 times the currently used value). Here we have plotted the two extreme cases of a Schwarzschild ($a=0$) and a maximally rotating BH ($a=1$). NGC~4051 is known to have a rapidly spinning black hole but not necessarily 1. In a recent study, \cite{Neeraj2023MNRAS} have found a lower limit on the BH spin i.e. $a>0.85$. So, for a slightly higher BH mass (2 times) and $0.85<a<1$, we expect our observed X--ray/UV lag to match the lag obtained by analytic equations. However, the X--ray/optical delay measured by \cite{Breedt2010MNRAS} still remains quite off from the model time--lags even for the corona height of $100~R_g$. For a low-mass AGN like NGC~4051, we do not expect such a larger time--lag between X--ray and optical bands. The large uncertainty in their lag values is a problem, however, if their measurements are reliable, it may indicate a drastic change or a different geometry of the corona. The plot in Fig.~\ref{lag_lambda} can put constraints on the current models regarding X--ray and UV/optical delays.


Previous lag measurements indicate that the X--ray corona height may be variable.
NGC~4051 is known to have short timescale flares in the X--ray band. From the time-resolved spectroscopy of NGC~4051 using simultaneous \xmm{} and \nustar{} data, \cite{Neeraj2023MNRAS} found that the reflection fraction decreases during X--ray flare due to an increase in the height of the corona. A similar analysis of another NLS1 I~Zwicky~1 (IZw~1), during and after a flaring period was performed by \cite{Wilkins2022MNRAS} and they also found a decrease in the reflection fraction and an increase in the 2-10 keV luminosity during the X--ray flare. They interpret that during the flare, the height of the corona changes (the corona extends vertically and/or radially) and is accelerated away from the black hole followed by a collapse into a compact region again after the flare. The dynamic behaviour of coronal height is also supported by the findings of \cite{Alston2020Nat} and \cite{Caballero_2020MNRAS} for IRAS~13224--3809. Due to the change in the coronal height, we also expect a change in the UV/X-ray time--lag for disc reprocessing of flaring X--rays.  Such dynamic variability (change in geometry and/or height) of the X--ray source in different segments of the light curve can lead to a low average UV/X--ray correlation strength \citep{Panagiotou2022ApJ}. In some cases, it may even lead to statistically insignificant lag values \citep{Mankatwit2023MNRAS, Chainakun2023MNRAS}.


\section{Summary}\label{sec5:conclusion}

We utilized $\sim4$ days long \astrosat{} data on NGC~4051, simultaneously observed in UV and X--ray bands, to investigate the UV/X--ray correlation and time--lag. The variations in the UV band are found to lag behind those in the X-ray band that supports the thermal disc reprocessing model. We also report the lag of NUV band photons relative to the FUV band with a high correlation ($\sim 0.9$). The FUV/NUV time--lag suggests that most of the NUV photons likely emanate from larger disc areas as compared to FUV. The comparative study of our X--ray/UV correlation results with the theoretical model for X--ray reverberation time--lag for different heights reveals that the SXT/NUV time--lag values are slightly higher than the model prediction (see Fig. \ref{lag_lambda}). This disagreement is possible due to the effect of X--ray ACF on CCF calculation in the model (explained in section \ref{sec:K21}). The uncertainty in other measured parameters such as BH mass, accretion rate and 2-10 keV luminosity may also lead to some deviation in the model time--lags.
We also noticed that the previously reported lags by other authors in X--ray/optical bands are much higher than the theoretical model time--lags. However, considering the large uncertainty on their lag values, if those measurements are reliable, it suggests that the geometry or height of the corona may have changed. The spectral analysis of NGC~4051 using \xmm{} and \nustar{} observations also supports the change in the height of the corona during flaring periods. 
The presence of short timescale flares in this source makes it even more intriguing for further investigation using long and good--quality data.

\section*{Acknowledgements}
 
This study utilized data from the \astrosat{} mission conducted by the Indian Space Research Organisation (ISRO), which is archived at the Indian Space Science Data Centre (ISSDC). The work also benefited from data obtained through the Soft X-ray Telescope (SXT) developed at TIFR, Mumbai. The UVIT project represents a collaborative effort involving IIA Bengaluru, IUCAA Pune, TIFR Mumbai, and various ISRO and CSA centres. The SXT POC at TIFR and UVIT POC at IIA in Bangalore played a crucial role in verifying and releasing the data via the ISSDC data archive. The UVIT data were processed using the CCDLAB pipeline. K. P. Singh thanks the Indian National Science Academy for support under the Senior Scientist programme. This research made use of APLpy, an open-source plotting package for Python.


\section*{Data Availability}
The \astrosat{} data are
available at \url{https://astrobrowse.issdc.gov.in/astro_archive/archive/Home.jsp}.



\bibliographystyle{mnras}
\bibliography{mybib} 



\appendix

\section{Spectral Analysis in X--ray band}\label{spec_fitting}

\begin{figure}[!ht]
\includegraphics[width=1\columnwidth]{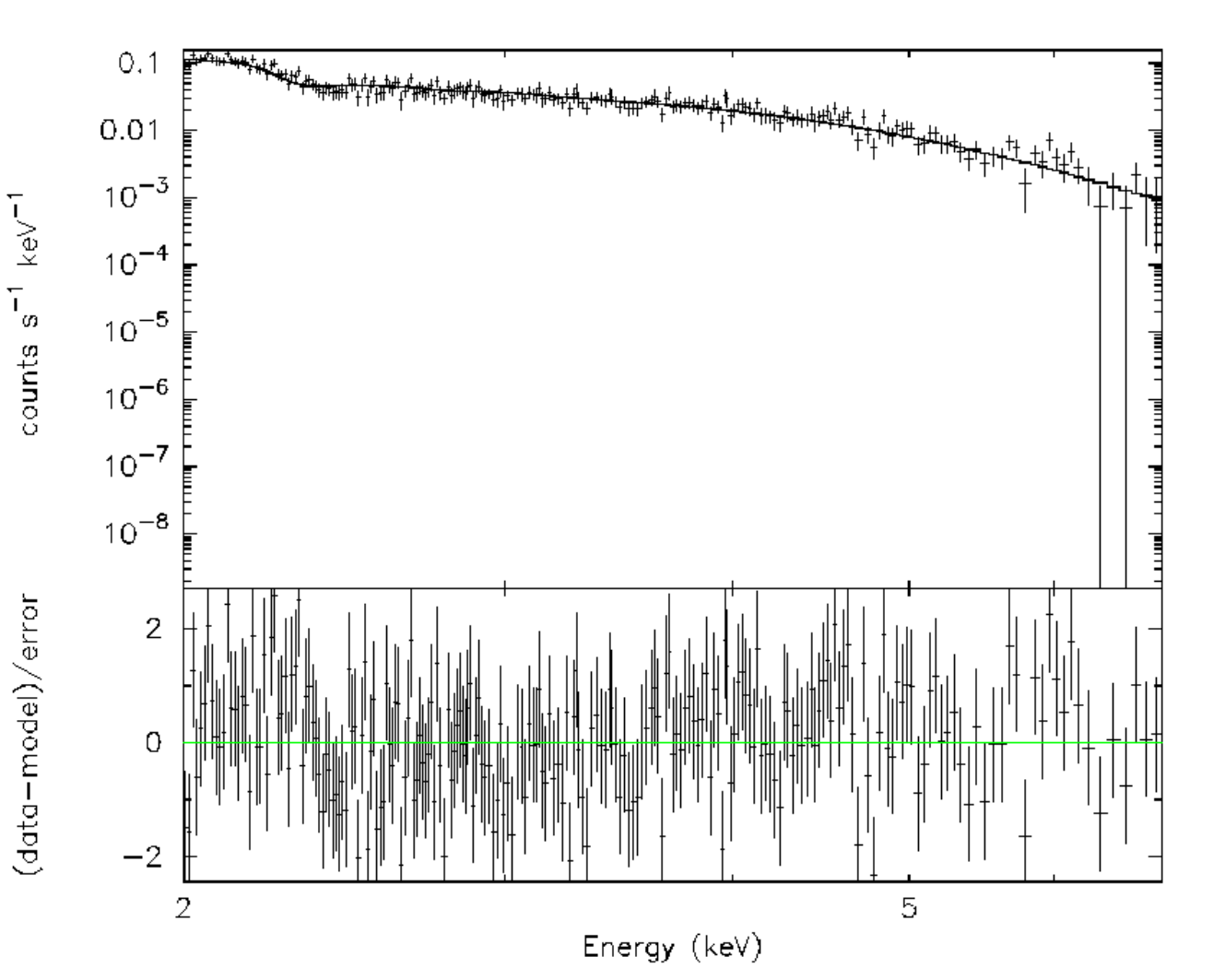}
\caption{The X--ray spectrum in 2-7 keV band, fitted with the model: \textit{TBabs$\times$zpowerlw}. }
\label{sxt_fit}
\end{figure}

We fitted the SXT spectrum in the 2-7 keV band using XSPEC (version: 12.12.1). With the simple model: \textit{TBabs$\times$zpowerlw} we obtained the statistically good fit with $\chi^2/d.o.f = 219.45/226$ (equivalent to 0.97). We do not see any significant feature in the residual (Fig. \ref{sxt_fit}). The best--fit models include the redshifted power-law \textit{`zpowerlw'} modified by the Galactic absorption \textit{`TBabs'}. The hydrogen column density representing 
the Galactic absorption along the line of sight held fixed to $N_H = 1.19 \times 10^{20} cm^{-2}$. The best--fit model yielded the power-law photon index $\Gamma = 2.01^{+0.08}_{-0.08}$. 

To estimate the flux in the 2-10 keV band, we utilized the convolution model `\textit{cflux}' in XSPEC. We multiply \textit{cflux} with \textit{zpowerlw} and fix the normalization parameter. We set the energy limits as $Emin=2$~keV and $Emax=10$~keV. This left only the $\Gamma$ and \textit{Flux} (in log$_{10}$ unit) as varying parameters in the fit.  Following the fit, we calculated the flux, $F_{2-10keV}= 2.05\times10^{-11}~erg~s^{-1}cm^{-2}$. Subsequently, using the redshift--independent distance, $D= 16.6 $~Mpc (\citealt{Yuan2021}), we computed the luminosity in 2-10 keV band,  $L_{X} = 6.79\times 10^{41}$ $erg~s^{-1}$.

\bsp	
\label{lastpage}
\end{document}